# Therapeutic Management of Sacro-coccygeal compression in a Green Iguana (*Iguana iguana*)

Senthilkumar. K[1]*, Ruchika Lakshmanan[2], Mohamed Shafiuazama[3] and Arun Prasad.A[4]

1Post graduate research Institute in Animal Sciences, Kattupakkam- 603 203, Tamil Nadu Veterinary and Animal Sciences University (TANUVAS), Tamil Nadu, India; 2- Department of wildlife science, Madras Veterinary College, Chennai-600 007, Tamil Nadu Veterinary and Animal Sciences University (TANUVAS), Tamil Nadu, India 3- Department of Veterinary Surgery, Madras Veterinary College, Chennai-600 007, Tamil Nadu Veterinary and Animal Sciences University (TANUVAS), Tamil Nadu, India and 4- Peripheral Veterinary Hospital, Madhavaram, Chennai-600051, , Tamil Nadu Veterinary and Animal Sciences University (TANUVAS), Tamil Nadu, India

*Correspondence: K.Senthilkumar, Post graduate research Institute in Animal Sciences, Kattupakkam- 603 203, Tamil Nadu Veterinary and Animal Sciences University (TANUVAS), Tamil Nadu, India;
E-mail: senthilkumar.k.wls@tanuvas.ac.in

## ABSTRACT

**Back ground:** In reptiles, especially during their early stage of life when reared under captive conditions, nutritional deficiencies are the most common problems reported in them affecting the spinal cord and the musculoskeletal system. More often green iguanas and snakes are very much vulnerable to it. **Case description:** A three and a half months old green iguana (*Iguana iguana*) was brought to the Avian and Exotic Pet Unit, Madras Veterinary College with the history of dragging hind limbs and not passed faeces for the past four days. **Findings, treatment and outcome:** On examination of hind limbs, it was found that perching reflex was absent and the abdomen was distended. It was tentatively diagnosed for hind limb paralysis but the radiological examination revealed that the iguana was suffering from Sacro-coccygeal compression. The condition could have been aggravated due to poor nutritional management also**.** In order to reduce the pain, the iguana was treated with Meloxicam orally @ 0.1 mg/kg for a week. In addition, it was successfully treated orally with multivitamin syrup and calcium syrup simultaneously for a month. The iguana was subjected to infrared therapy of 3 minutes for a period of one week. The reptile showed complete recovery without dragging its hind limb after 2 months. Conclusion: It could be concluded that prompt intervention and dietary management may be necessary for musculo-skeletal disorders in Iguana. .

**Key Words:** - Green iguana, proliferative osteo arthropathy, Metabolic Bone Disease

## Introduction

In reptiles, especially during their early stage of life when reared under captive conditions, nutritional deficiencies are the most common problems reported in them affecting the spinal cord and the musculoskeletal system. More often green iguanas and snakes are very much vulnerable to it. Improper husbandry and diet, traumatic injury, poor nutrition (calcium and vitamin D deficiency) are the main etiological factors associated with metabolic bone disease *(Klaphake, 2010) and* this, in turn, results in Proliferative Spinal Osteo Arthropathy *( Mayer and* Donnelly, 2013*). This* is a condition in which there is anomalous displacement or distortion of spinal column characterized by segment fusion of affected vertebrae by foci of irregular proliferative bone.

## Case Description

A three and half months old Green Iguana *(Iguana iguana)* was brought to Avian and Exotic Pet Unit of Madras Veterinary College with the history that the iguana jumped from a height and since then it is dragging its limbs, constipating for past four days and the abdomen is distended.

On clinical examination, it was recorded that the perching reflex was absent in the hind limbs, there was pain on palpation of the spinal region and there was dragging of hind limbs. Tentatively, it was diagnosed for hind limb paralysis. The iguana was subjected to radiological examination (Zotti *et al.,* 2004) by placing it on the X-ray plate (Fig.1). The radiological examination (Fig.2 & Fig.3) revealed that the iguana is suffering from Sacro-coccygeal compression.

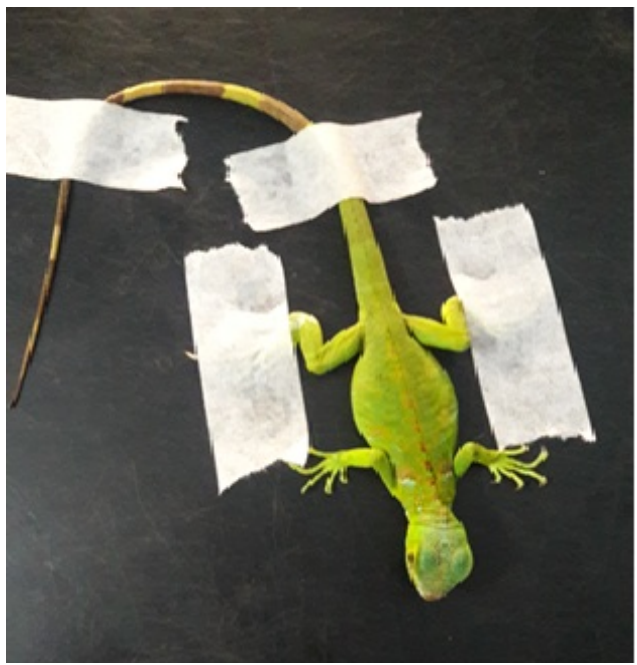

Fig.1 Radiological examination of Iguana

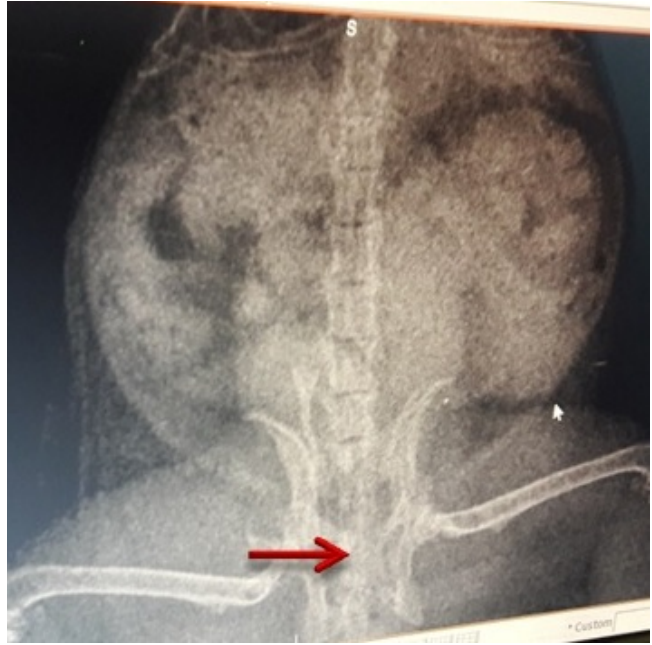

Fig. 2: Sacro-coccygeal compression of vertebral bone (arrow) in the pelvic area (Ventral View).

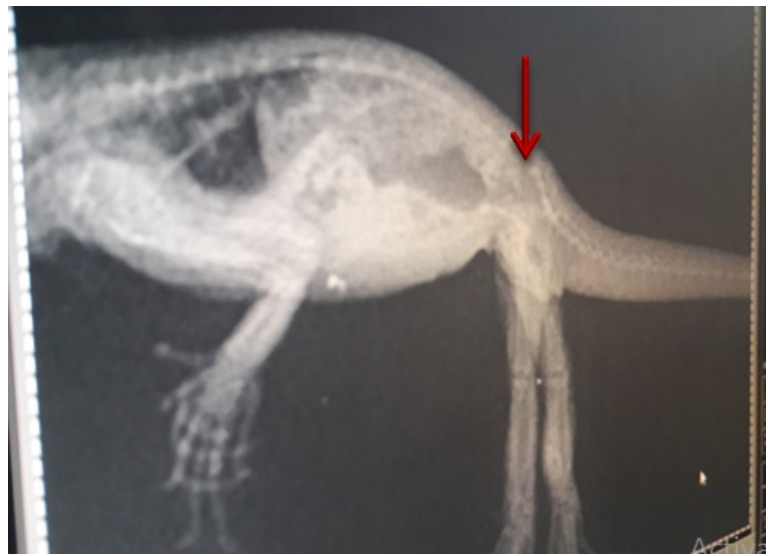

Fig. 3 Sacro-coccygeal compression of vertebral bone (arrow) in the pelvic area (lateral View).

When the distended abdomen was massaged, the iguana passed faeces. In order to alleviate pain Tab. Meloxicam was given orally @0.1 mg/kg b.wt., *( James Carpenter, 2013)*. Multivitamin syrup and calcium syrup was prescribed for a month*,* orally. It was subjected to infrared therapy of 3minutes for a period of a week (Fig.4). Proper nutrition, feeding, housing and management advice was given *(*Divers & Mader, 2005*)*. There were no complications during the follow up period of 8 months and the iguana was healthy

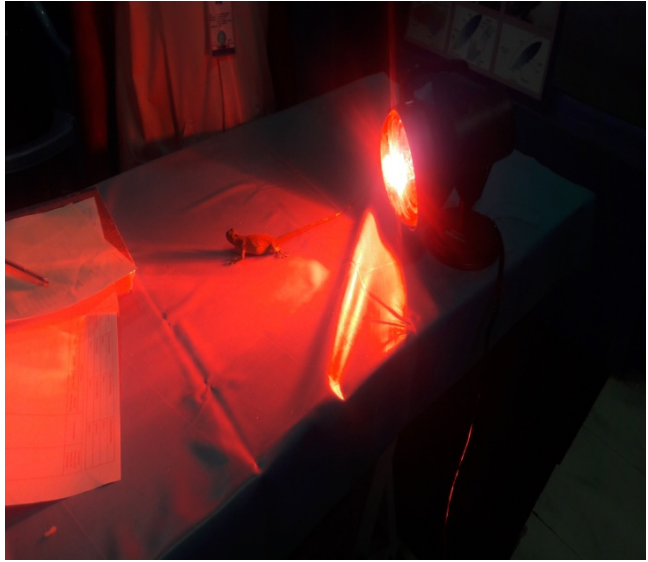

Fig.4 Infra-red therapy to the Iguana

**Discussion**

The sacro-coccygeal compression in this iguana is mainly due to weak vertebral bones and also a juvenile, which will have friable bones (Zotti *et al.,* 2004). In case of green iguanas the diet is very essential. In nature, iguanas are herbivores strictly (Govende *et al.,* 2012), hence herbivorous diet routine should be followed regularly under captive condition. A huge number of captive iguanas were suffering from this condition because they are improperly fed by the owners as they are having misconception that they are being coming under lizard category, they will fed with insects, which results in indigestion and nutritional deficiencies.

They should be generally fed with high fibre diet which includes leaves, green leafy vegetables, mushrooms, turnip green, and bell peppers (Troyer, 1984). The diet should have balanced calcium and phosphorous content and low oxalates which favours calcium absorption (Donoghue,1994). Spinach should be reduced as it is rich in oxalates and fruits should be minimised as they dilute beneficiary nutrients. Strawberries, bananas and fruits rich in calcium are recommended in its diet.

Basking in sunlight is a very essential practice that has to be practiced every morning because light is essential for the metabolism and vitamin D3 (Mans and Braun, 2014).The cages for the baby iguana should be of size of 40-gallon terrarium with a screen lid will accommodate a young **iguana** to prevent self-inflicted injuries.